\RequirePackage{fix-cm}
\documentclass[smallextended]{svjour3}
\smartqed

\usepackage{hhline,amsgen,bm,graphicx,amsfonts,lineno,amssymb,float,cite,bbold,setstack,amsgen,mathrsfs,microtype}
\usepackage[font={footnotesize,it}]{caption}
\usepackage[caption = false]{subfig}
\usepackage{afterpage,lscape}

\def\makeheadbox{\relax}

\begin{document}

\title{A new model for strange stars}

\author{Debabrata~Deb, Sourav~Roy~Chowdhury, Saibal~Ray and Farook~Rahaman}

\institute{Debabrata~Deb$^{1}$, Sourav~Roy~Chowdhury$^1$, Saibal~Ray$^2$ and Farook~Rahaman$^3$\\ \\
              1. Department of Physics, Indian Institute of Engineering Science and Technology, Shibpur, 
                 Howrah, West Bengal, 711103, India. \\
              2. Department of Physics, Government College of Engineering and Ceramic Technology, Kolkata 700010, 
                 West Bengal, India.\\
              3. Department of Mathematics, Jadavpur University, Kolkata 700032, West Bengal, India.\\ \\
              \email{ddeb.rs2016@physics.iiests.ac.in,~sourav.rs2016@physics.iiests.ac.in,~saibal@associates.iucaa.in, rahaman@associates.iucaa.in}}

\date{}

\maketitle

\begin{abstract}
In the present work, we attempt to find a new class of solutions for 
the spherically symmetric perfect fluid sphere by employing the Homotopy 
Perturbation Method (HPM), a new tool via which the mass polynomial
function facilitates to tackle the Einstein field equations. A set of interior solutions found 
on the basis of the simplest MIT bag model equation of state (EOS) in the form 
$p=\frac{1}{3}(\rho-4B)$ where $B$ is the bag constant. The proposed interior 
metric for the stellar system is consistent with the exterior Schwarzschild 
spacetime on the boundary. In addition, we also conduct a detailed study on different tests, 
viz. the energy conditions, TOV equation, adiabatic index, Buchdahl limit, etc., 
to verify the physical validity of the proposed model. The numerical 
value of the used parameters is predicted for different strange star candidates, 
for different chosen values of the bag constant. In a nutshell, by exploiting 
HPM technique first time ever in the field of relativistic astrophysics, we have 
predicted in the present literature  a singularity-free and stable stellar model 
which is suitable to describe ultra-dense objects, like strange (quark) stars.

\end{abstract}
\vspace{1.0cm}
Keywords: General relativity . homotopy perturbation method . strange stars

\newpage

\section{Introduction}\label{intro}
Several scientists~\cite{Bodmer1971,Terazawa1979,Witten1984} have
pointed out that the matter made of $u$, $d$ and $s$
quarks, and some electrons (to ensure the charge neutrality) known
as the strange quark matter, may be more stable than the ordinary
nuclear matter. The strange stars, named after the {\textit{strange quark}},
are therefore composed of strange matter. The interesting
difference between the strange stars and neutron stars is that the
former one can vary in size from roughly 0 to 12 km 
whereas neutron stars are mostly of the radius $>12$ km. 

Quark matter is self-bounded by the forces of quantum
chromodynamics (QCD). Therefore, like neutron stars, which are
gravitationally bounded, the stability of a strange star is 
independent of the gravity~\cite{Haensel1986}. However, this statement
is not true in general, e.g. a strange star, having a central energy 
density slightly above the maximum mass limit is not stable and
due to the gravitational force, it needs to collapse to a black
hole. This stability threshold depends on the underlying
gravitational interaction and differs between alternative theories
of the gravity. Essentially the degeneracy pressure of the nucleons 
within a neutron star is balanced by the gravitational force, and hence
for the star to be stable its mass must be greater than a certain value.

In 1916 first time ever Karl Schwarzschild~\cite{Schwarzschild1916} presented 
an exact solution to the Einstein field equations for a spherically symmetric 
isotropic system. Later in 1939 Oppenheimer and Volkoff~\cite{Oppenheimer1939} 
introduced the equation of hydrostatic equilibrium for isotropic spherically symmetric stellar configuration.
In the same year, Tolman~\cite{Tolman1939} presented seven solutions of the 
Einstein field equations. Delgaty and Lake~\cite{Lake1998} in their pioneering 
work showed that for isolated, static and spherically symmetric perfect fluid 
stellar system only 16 solutions of Einstein's field equations out of available 
127 solutions are physically acceptable. It is worth mentioning that in this 
line several scientists~\cite{Finch1998,Nilsson2000,Visser2002,Lake2003,Martin2004a,Boonserm2005} 
attempted to produce a physically acceptable solution of the Einstein field equation 
for the isotropic spherically symmetric stellar system. 

Solving  analytically the non-linear equations it has always been a
challenge to the astrophysicists. The Homotopy Perturbation Method (HPM)
is a powerful as well as very simple tool to solve this kind of
equations with the least number of assumptions. To solve the differential and 
integral equations He~\cite{He1998a,He1998b} first proposed the semi-analytical HPM 
technique in 1998 and later on improved it further~\cite{He1999,He2000}. Immediately, 
it drew attention to many researchers~\cite{Mallil2000,Hussein2000,Hillermeier2001,Cadou2001,Mokhtari2001,Jegen2001,He2004} 
for solving non-homogeneous as well as non-linear partial differential equations. The important 
study by Cveticanin~\cite{Cveticanin2006} has revealed that for a large range of nonlinear 
problems in the applied and fundamental science it is very advantageous to use the HPM 
technique which provides an analytical approximate solution. As the obtained solutions 
by the technique of HPM appear as the rapidly converging infinite series, hence 
it is enough to limit the calculations for the first few orders. It is worth mentioning 
that recently in both the field of cosmology and astrophysics several authors~\cite{Zare2012,Shchigolev2014,Shchigolev2015,Shchigolev2015a} have successfully used the HPM technique in their studies. Interestingly, unlike other usual perturbation 
techniques, any restrictive assumption, linearization or discretization is not essential 
for HPM to obtain a simple and effective solution to the equations to be solved.

Using the MIT bag model, Rahaman et al.~\cite{Rahaman2014} have
obtained a deterministic model of strange stars, where they
considered a mass polynomial and analyzed all the physical
properties. However, they were unable to obtain the physical
properties of the model up to 6 km from the center of the
system. We would also refer to the work of Rahaman et al.~\cite{Rahaman2015}
where the HPM has been employed for a spherically symmetric system of radiating star
which suffers from stability condition. Again, Aziz et al.~\cite{Aziz2016} in 
their study used the HPM to describe a compact stellar system, but they failed to 
provide a singularity-free stable stellar model. In this context, it is worth 
mentioning that the present investigation is the first study where the HPM technique 
has been employed successfully to provide a singularity-free as well as stable spherically 
symmetric stellar system and thus offers a possibility to open up a new arena 
in the studies of relativistic astrophysics. 

In the present work, we have developed an expression of mass (in the polynomial form) 
as a function of the radial coordinate $r$ which is appropriate for the strange stars. 
However, it is not assumed arbitrarily, rather we have computed this by the help of HPM. 
Further, we have substituted that expression of mass to solve the Einstein field equations 
by using the MIT bag EOS in the form $p=\frac{1}{3}(\rho-4B)$. The entire solutions set 
thus obtained provides a stable model of ultra-dense compact stars. 

The outline of our investigation is as follows: In Sec.~\ref{EOS} we discuss the EOS for the quark stars 
and show the basic formalism of the HPM in Sec.~\ref{sec3}. To calculate the expression for the
 mass function of the system in Sec.~\ref{calmass}, firstly, we set up the basic and 
essential stellar structure equations ~\ref{MEP}, and then applied the HPM technique~\ref{AHPM}.
Sec.~\ref{solEFE} deals with the solution of the Einstein field equations for 
different physical parameters, viz. the pressure and energy density.
We have discussed and explored several physical features in Sec.~\ref{physical}
and a comparative study has been conducted in Sec.~\ref{com_st} for the validity
of the data set of the present model with the existing
strange stars available in the literature~\cite{Rawls2011,Guver2010a,Freire2011,Guver2010b,Demorest2010}.
In the last Sec.~\ref{discsn} we remark on some of the salient features of the present model.

\section{The MIT bag equation of state}\label{EOS}
Considering the three flavors of quarks ($u$, $d$ and $s$) as non-interacting, 
i.e. zero strong coupling constant and confined in a bag, the simplest and
linear form of the EOS can be written as
\begin{equation}
{p}+{B}={\sum_f}{p^f},\label{EOS1}
\end{equation}
where the external bag pressure $B$ counterbalanced the sum 
of the individual pressures $p^f$ of all the quarks. 
The masses of the quark matter are much higher than the chemical potentials 
involved in the system $(\simeq~300~MeV)$. Also, we exclude the effects of leptons 
in the system since in the present case the leptons are not required 
to electrically neutralize the phase~\cite{Farhi1984}.

The deconfined quarks inside the bag have the total energy density $\rho$ given by
\begin{equation}
\rho={\sum_f}{{\rho}^f}+B, \label{EOS2}
\end{equation}
where ${{\rho}^f}=3{p^f}$ is energy density of the individual quarks.

Using Eqs.~(\ref{EOS1}) and (\ref{EOS2}) the EOS of the matter distribution adopts the simple form as follows
\begin{equation}
p=\frac{1}{3}(\rho-4\,B).\label{EOS3}
\end{equation} 

Eq.~(\ref{EOS3}) is featuring the well-known MIT bag EOS to describe strange quark stars. 
The successful use of this EOS can be found in the recent several works~\cite{1,2,3,4,5,6,7,8}. 
However, Kalam et al. in their work~\cite{Kalam2002} showed that a wide range of values of the 
bag constant are allowed which are well supported by the recent CERN-SPS and RHIC data~\cite{Burgio2002}. 
Therefore, in the present study, following the proposals of Farhi and Jaffe~\cite{Farhi1984} 
and Alcock et al. ~\cite{Alcock1986} we choose higher values of bag constant arbitrarily as $83~MeV/{fm}^3$~\cite{Rahaman2014}, $100~MeV/{fm}^3$, and $120~MeV/{fm}^3$.

\section{Basic formalism of the Homotopy Perturbation Method}\label{sec3}
In order to demonstrate the basic formalism of HPM for solving 
nonlinear differential equations, let us consider a general nonlinear differential equation given as
\begin{eqnarray}
L \left( u \right) +N \left( u \right) =f \left( r,t \right);~~r\in\Omega,\label{3.1}
\end{eqnarray}
with the boundary condition 
\begin{eqnarray}
B\left(u,\frac{\partial u}{\partial n}\right)=0;~~r\in\gamma,\label{3.2}
\end{eqnarray}
where $L$ is a linear operator, $N$ is a non-linear operator, $f\left(r,t\right)$ 
is a known analytical function, $B$ is the boundary operator and $\gamma$ 
is the boundary of the domain $\Omega$.

By using the homotopy method, one can construct a homotopy
\begin{eqnarray}
v\left(r,\mathfrak{p}\right): \gamma \times \left[0,1\right]\rightarrow R,\label{3.3}
\end{eqnarray}
which satisfies~\cite{He2000}
\begin{eqnarray}\label{3.4}
 H \left( v,\mathfrak{p} \right) = \left( 1-\mathfrak{p} \right) \left[L \left( v \right) -L
 \left( u_{{0}} \right) \right]+\mathfrak{p}\left[L \left( v \right) +N \left( v \right) -f
 \left( r,t \right) \right]=0,\\ \label{3.5}
 H \left( v,\mathfrak{p} \right)=L \left( v \right)-L \left( u_{{0}} \right) + \mathfrak{p}\left[L \left( u_{{0}} \right) +N \left( v \right) -f
 \left( r,t \right) \right]=0,
\end{eqnarray}
where $\mathfrak{p}\in\left[0,1\right]$ is an embedding parameter 
and $u_{{0}}$ is the initial approximation which is nothing but the initial value 
of the unknown $u$. Here
\begin{eqnarray}\label{3.6}
 H \left( v,0 \right)=L \left( v \right)-L \left( u_{{0}} \right)=0,\\ \label{3.7}
 H \left( v,1 \right)=L \left( v \right)+N \left( v \right)-f \left( r,t \right)=0.
\end{eqnarray}

The changing process of $\mathfrak{p}$ from $0$ to $1$ is nothing 
but $v\left(r,\mathfrak{p}\right)$ changes from $u_0$ to $u\left(r\right)$. 
This is known as the deformation of homotopy and $L \left( v \right)-L \left( u_{{0}} \right)$ 
and $L \left( v \right)+N \left( v \right)-f \left( r,t \right)$ are homotopic.

The solution of Eq.~\ref{3.7} can be expressed as a power series of $\mathfrak{p}$ and 
is given by
\begin{eqnarray}
v=u_{{0}}+\mathfrak{p}v_{{1}}+{\mathfrak{p}}^{2}v_{{2}}+....\label{3.8}
\end{eqnarray}

By the choice of $\mathfrak{p}\rightarrow1$, Eq.~(\ref{3.5}) reduces to 
Eq.~(\ref{3.1}). Again, Eq.~(\ref{3.8}) turns into the approximate solution 
of Eq.~(\ref{3.1}) and can be written as
\begin{eqnarray}
\lim_{\mathfrak{p}\to 1} v=u_{{0}}+v_{{1}}+v_{{2}}+....\label{3.9}.
\end{eqnarray}

The series in Eq.~(\ref{3.9}) is a convergent series for most of the cases. 
However, convergence rate depends on the non-linear operator $N \left( v \right)$.

\section{Calculation of mass of the spherical system}\label{calmass}

\subsection{Stellar structure equations}\label{MEP}
To investigate the spherically symmetric compact stellar model of perfect
fluid we are using the equation of state (EOS) in the following form
\begin{equation}
p=\frac{1}{3}(\rho-4B), \label{eq1}
\end{equation}
where $p$ is the pressure and $\rho$ is the energy density of the matter distribution inside the compact star.

We consider the interior spacetime metric of the spherical
symmetric stellar system as (in natural units $G=c=h=k=1$)
\begin{equation}
\mathrm{ds}^{2}=-g_{tt}(r)\mathrm{dt}^2+
g_{rr}(r)\mathrm{dr}^2+  r^{2}
(\mathrm{d\theta}^2+\sin^{2}\theta\mathrm{d\phi}^2), \label{eq4}
\end{equation}
where $m(r)$ is the mass distribution of the system. The metric potentials, 
i.e., $g_{tt}$ and $g_{rr}$ are the function of the radial component $r$ only.

The general energy-momentum tensor for the spherically symmetric
perfect fluid system is as follows
\begin{equation}
T_\nu^\mu= (\rho + p)u^{\mu}u_{\nu} + p g^{\mu}_{\nu},\label{eq5}
\end{equation}
with $u^{\mu}u_{\mu}=1$. Here the vector $u^{\mu}$ is the fluid 4-velocity of the local
rest frame.

The Einstein field equations for the metric (\ref{eq4}) and the
matter distribution given in Eq. (\ref{eq1}) can be written as
\begin{eqnarray}\label{eqE1}
&\qquad  \frac{1}{g_{{{ \mathit{rr}}}}}\left(\frac{1}{r} { \frac{{\mathrm d} \ln  \left( g_{{{\mathit{rr}}}}  \right)}{{\mathrm d}r} }-\frac{1}{r^2}\right)+\frac{1}{r^2}=8\pi\rho, \\ \label{eqE2}
&\qquad \frac{1}{g_{{{\mathit rr}}}}\left(\frac{1}{r^2}+\frac{1}{r} { \frac{{\mathrm d} \ln  \left( g_{{{\mathit tt}}}  \right)}{{\mathrm d}r} } \right)-\frac{1}{r^2}= {8\pi}p, \\ \label{eqE3}
&\qquad\hspace{-1.10cm} \frac{1}{2 {g_{rr}}}\left[\frac{1}{2}\left(\frac{{\mathrm d} \ln  \left( g_{{{\mathit tt}}}  \right)}{{\mathrm d}r}\right)^2+{ \frac{{\mathrm d^2} \ln  \left( g_{{{\mathit tt}}}  \right)}{{\mathrm d}r^2} }-\frac{1}{2}{ \frac{{\mathrm d} \ln  \left( g_{{{\mathit rr}}}  \right)}{{\mathrm d}r} }{ \frac{{\mathrm d} \ln  \left( g_{{{\mathit tt}}}  \right)}{{\mathrm d}r} }+\frac{1}{r}\left({ \frac{{\mathrm d} \ln  \left( g_{{{\mathit tt}}}  \right)}{{\mathrm d}r} }-{ \frac{{\mathrm d} \ln  \left( g_{{{\mathit rr}}}  \right)}{{\mathrm d}r} }\right)\right] = {8\pi}p. \nonumber\\
\end{eqnarray}

For the present spherically symmetric system the mass function can be defined as
\begin{eqnarray}
m \left( r \right) =4\pi\int_{0}^{r}\!\rho \left( r \right) {r}^{2}{dr}.\label{masseq}
\end{eqnarray}

We assume that the exterior spacetime is governed by the well known Schwarzschild metric given as
\begin{equation}
\mathrm{ds}^{2}=-\left(1-\frac{2M}{R}\right) \mathrm{dt}^2+\left(1-\frac{2M}{R}\right)^{-1}\mathrm{dr}^2+  r^{2}
({\mathrm{d}{\theta}}^2+\sin^{2}\theta {\mathrm{d}\phi}^2). \label{exterior}
\end{equation}

Using Eqs.~(\ref{eqE1}), (\ref{masseq}) and (\ref{exterior}) we obtain the metric component $g_{rr}$ as follows
\begin{eqnarray}
g_{rr}=\left(1-\frac{2m}{r}\right)^{-1}.\label{MC1}
\end{eqnarray}

Now, the stellar structure equations are essential to describe the spherically symmetric isotropic system and are given by
\begin{eqnarray}\label{eq6}
&\qquad {\frac {{\mathrm d} m}{{\mathrm d}r}}=4\pi{r}^{2}\rho, \\ \label{eq6.1}
&\qquad \frac{{\mathrm d} \ln  \left( g_{{{\mathit rr}}}  \right)}{{\mathrm d}r}=-\frac{2}{\left(\rho+p\right)}{\frac {{\mathrm d} p}{{\mathrm d}r}}.
\end{eqnarray}

Here Eq.~(\ref{eq6.1}) represents the conservation of the energy-momentum tensor, $T_\nu^\mu$ and also known as the Tolman-Oppenheimer-Volkoff (TOV) equation~\cite{Tolman1939,Oppenheimer1939}. Again, substituting Eqs.~(\ref{eqE2}) and (\ref{MC1}) into Eq.~(\ref{eq6.1}) we have the novel hydrodynamic equation given as
\begin{eqnarray}\label{eq6.2}
{\frac {{\mathrm d} p}{{\mathrm d}r}}=-\left(\rho+p\right)\frac{\left(4{\pi}rp+\frac{m}{r^2}\right)}{\left(1-\frac{2m}{r}\right)}.
\end{eqnarray}   

By substituting Eqs.~(\ref{eq1}) and ({\ref{eq6}) into Eq.~(\ref{eq6.2}) we obtain
\begin{eqnarray}
& \qquad\hspace{-1cm} m^{\prime\prime}+{\frac {256}{3}}{{B}^{2}{\pi }^{2}{r}^{3}}-{\frac {80}{3}}{B\pi r m^{{\prime}}} -16\,B\pi m\nonumber \\
& \qquad\hspace{3cm}-{\frac {2 m^{{\prime\prime}}m}{r}}+\frac{4}{3}{\frac {{m^{{\prime}}}^{2}}{r}}+{\frac {8 m m^{{\prime}}}{{r}^{2}}}-{\frac {2 m^{{\prime}}}{r}}=0,
\label{eq10}
\end{eqnarray}
where `$\prime$' denotes the derivation with respect to $`r$'. The above non-linear differential equation is nothing but the TOV equation in terms of the mass function $m(r)$ only. By solving this Eq. (\ref{eq10}) one can obtain the expression for the mass profile of the stellar system. Now to solve the highly non-linear differential Eq. (\ref{eq10}) we shall apply the HPM technique in the following subsection.

\subsection{Application of the HPM}\label{AHPM}
For a spherical symmetric stellar system, an initial expression of mass 
can be chosen as $m(r)=a r^3$, where $a$ is a constant. We have 
already mentioned in Sec.~\ref{sec3} that using the formalism 
HPM as provided by He~\cite{He2000} we have calculated the expression of $m$. 
Hence, the Homotopy for the non-linear differential equation~(\ref{eq10}) takes form as
{{\begin{eqnarray}\label{4.1.1}
& \qquad\hspace{-0.4cm} m^{\prime\prime}-{{m}_0}^{\prime\prime}+\mathfrak{p}\Big[{{m}_0}^{\prime\prime}+{\frac {256\,{B}^{2}{\pi }^{2}{r}^{3}}{3}}-{\frac {80\,B\pi \,r m^{{\prime}}}{3}} -16\,B\pi \,m\nonumber \\
& \qquad\hspace{4cm}-{\frac {2 m^{{\prime\prime}}m}{r}}+\frac{4}{3}{\frac {{m^{{\prime}}}^{2}}{r}}+{\frac {8 m m^{{\prime}}}{{r}^{2}}}-{\frac {2 m^{{\prime}}}{r}}\Big],
\end{eqnarray}}}
where $\mathfrak{p}$ is the embedding parameter such as $\mathfrak{p}\in\left[0,1\right]$. 

To find out the expression for $m$, we consider the general solution of $m$ as follows:
\begin{equation}
m=(m_0+\mathfrak{p}^{1}m_1+\mathfrak{p}^{2}m_2+....) .\label{eq12}
\end{equation}

As, mentioned above the chosen initial condition is given as
\begin{eqnarray}\label{Eqini}
{m_0}\left(r\right)=ar^3.
\end{eqnarray}

However, the initial boundary condition can be chosen as
\begin{eqnarray}\label{4.1.2}
{m_0}\left(0\right)={m_0}^{\prime}\left(0\right)=0, \\ \label{4.1.3}
{m_i}\left(0\right)={m_i}^{\prime}\left(0\right)=0,
\end{eqnarray}
where $i>1$. 

Now substituting Eq.~(\ref{eq12}) into Eq.~(\ref{4.1.1}) we have 
{\small{\begin{eqnarray}\label{4.1.4}
& \qquad \mathfrak{p}^0:~~~~{{m_0}^{\prime\prime}}-{{m_0}^{\prime\prime}}=0\\ \label{4.1.5}
& \qquad \mathfrak{p}^1:~~~~{{m_1}^{\prime\prime}}+{{m_0}^{\prime\prime}}+{\frac {256\,{B}^{2}{\pi }^{2}{r}^{3}}{3}}-{\frac {80\,B\pi \,r {m_0}^{{\prime}}}{3}}-16\,B\pi \,{m_0}-{\frac {2 {m_0}^{{\prime\prime}}{m_0}}{r}}\nonumber \\
& \qquad\hspace{5.5cm} +\frac{4}{3}{\frac {{{m_0}^{{\prime}}}^{2}}{r}}+{\frac {8 {m_0} {m_0}^{{\prime}}}{{r}^{2}}}
-{\frac {2 {m_0}^{{\prime}}}{r}}=0\\\label{4.1.6}
& \qquad \mathfrak{p}^2:~~~~{{m_2}^{\prime\prime}}-{\frac {80\,B\pi \,r {m_1}^{{\prime}}}{3}}-16\,B\pi \,{m_1}-{\frac {2 {m_1}^{{\prime\prime}}{m_0}}{r}}--{\frac {2 {m_0}^{{\prime\prime}}{m_1}}{r}}\nonumber \\
& \qquad\hspace{3.5cm} +\frac{8}{3}{\frac {{{m_0}^{{\prime}}}{{m_1}^{{\prime}}}}{r}}+{\frac {8 {m_0} {m_1}^{{\prime}}}{{r}^{2}}}+{\frac {8 {m_1} {m_0}^{{\prime}}}{{r}^{2}}}-{\frac {2 {m_0}^{{\prime}}}{r}}=0.
\end{eqnarray}}}

By using the set of linear equations~(\ref{4.1.4})-(\ref{4.1.5}), chosen initial condition~(\ref{Eqini}) 
and the boundary conditions~(\ref{4.1.2}) and (\ref{4.1.3}), 
we get their solutions as
\begin{eqnarray}\label{EQsolu1}
&\qquad m_{{0}}\left( r \right) =a{r}^{3},\\\label{EQsolu2}
&\qquad m_{{1}}\left( r \right) =-\left( {\frac {256\,{B}^{2}{\pi }^{2
}}{3}}-96\,aB\pi +24\,{a}^{2} \right) \frac{{r}^{5}}{20},\\\label{EQsolu3}
&\qquad m_{{2}}\left( r \right) = \left( {\frac {256\,{B}^{2}{\pi }^{2}}{3}}-
96\,aB\pi +24\,{a}^{2} \right)  \left( -{\frac {8\,B\pi \,{r}^{7}}{45}
}+{\frac {13\,a{r}^{7}}{210}}-\frac{{r}^{5}}{40} \right).
\end{eqnarray}

\begin{figure}[h]
\centering
\includegraphics[scale=.3]{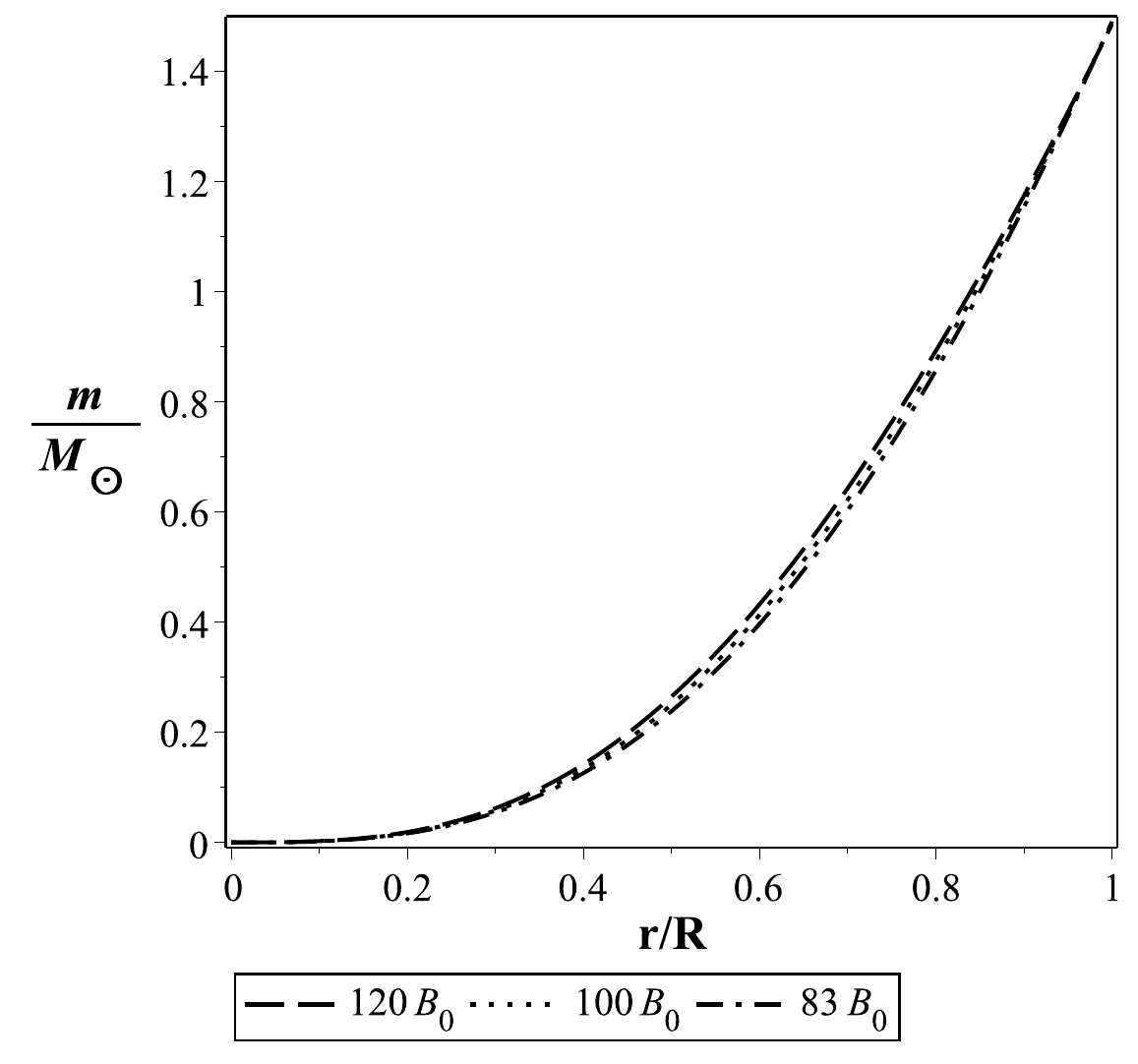}
\caption{Variation of the mass function $m$, normalized to the solar mass $M_{\odot}$ against the radial distance $r/R$ for the strange star candidate $Cen~X-3$. Here $B_0=1~MeV/{fm}^3$ }
\label{figmass}
\end{figure}

Here our calculation is intentionally limited to the minimum degree of approximation. Hence applying the HPM method by using the solutions ({\ref{EQsolu1}})-({\ref{EQsolu3}}) we have the final solution of Eq.~(\ref{4.1.1}) as
{\footnotesize{\begin{eqnarray}\label{eq13}
& \qquad\hspace{-2cm} m=lim_{\mathfrak{p}\to 1}(m_0+\mathfrak{p}^{1}m_1+\mathfrak{p}^{2}m_2+....) \nonumber \\
& \qquad\hspace{0.8cm} =a{r}^{3}+ \left( {\frac {256\,{B}^{2}{\pi }^{2}}{3}}-96\,aB\pi +24\,{a
}^{2} \right)  \left( {\frac {13\,a}{210}}-{\frac {8\,B\pi }{45}}
 \right) {r}^{7}\nonumber \\
& \qquad\hspace{5cm}- \left( {\frac {32\,{B}^{2}{\pi }^{2}}{5}}-{\frac {36
\,aB\pi }{5}}+\frac{9}{5}\,{a}^{2} \right) {r}^{5},
\end{eqnarray}}}
where for the sake of simplicity, we have limited our solution up to third order of approximation. The variation of the mass function, $m$ with the radial coordinate $r/R$ for the different chosen values of $B$ has been featured in Fig~\ref{figmass}, which shows the regularity of the achieved solution as $m=0$ at the center.

\section{The solution of Einstein's field equations}\label{solEFE}
Now substituting Eq. (\ref{eq13}) in Eq. (\ref{eqE1}), we get the energy density of the system as
\begin{equation}
\rho={\frac {1}{4\pi \,{r}^{2}} \left[ 3\,a{r}^{2}+7\,\rho_{{1}}
 \left( {\frac {13\,a}{210}}-{\frac {8\,B\pi }{45}} \right) {r}^{6}-\frac{3}{8}\rho_{{1}}{r}^{4} \right]},\label{eq15}
\end{equation}
where $\rho_{{1}}={\frac {256\,{B}^{2}{\pi }^{2}}{3}}-96\,aB\pi +24\,{a}^{2}$. The behaviour of this energy density are shown in Fig. \ref{Fig1}.

\begin{figure}[h]
\centering
\includegraphics[scale=.3]{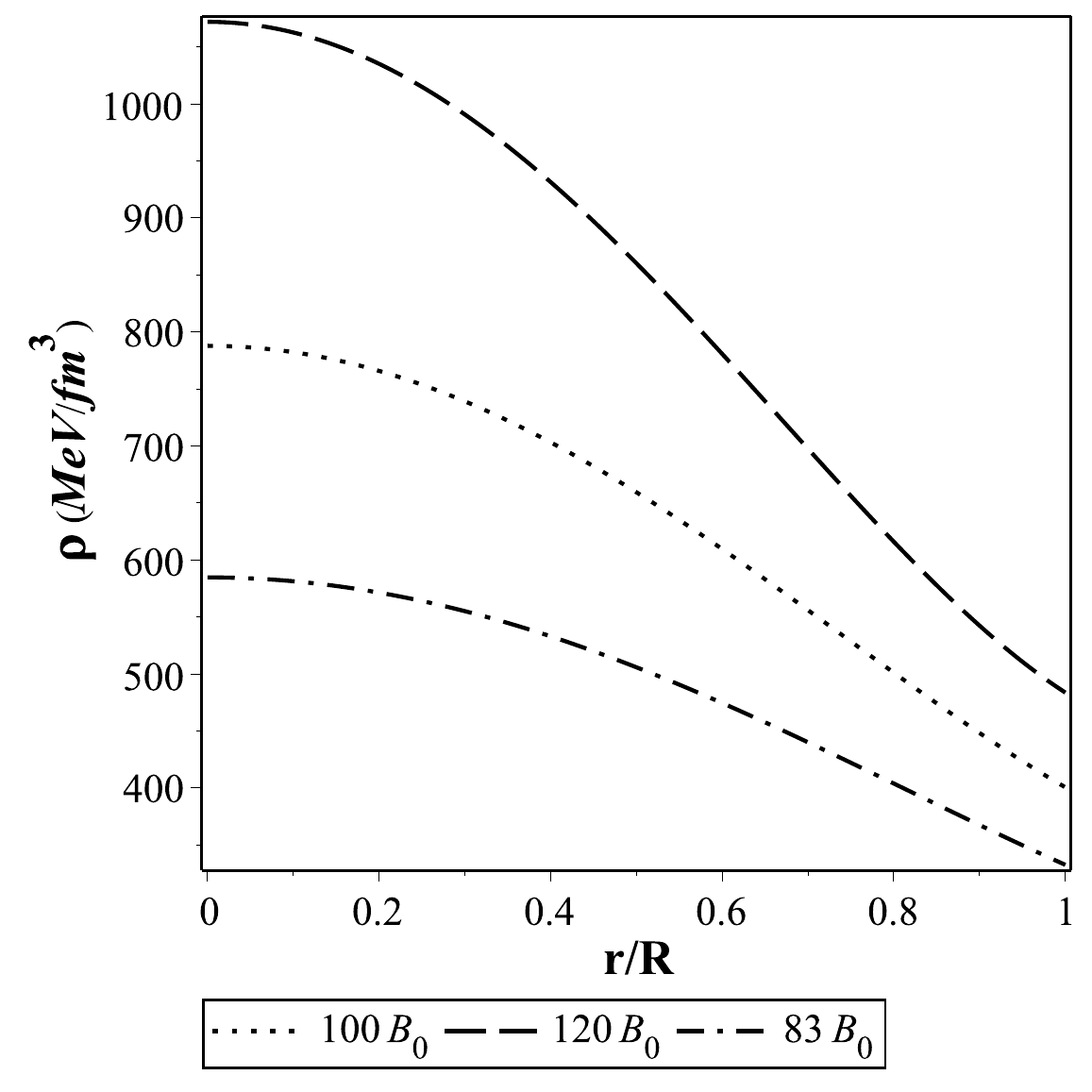}
\caption{Variation of the energy density as a function of the radial distance $r/R$ for the strange star $Cen~X-3$}
\label{Fig1}
\end{figure}

From Eqs. (\ref{eq1}) and (\ref{eq15}) one get
\begin{equation}
p = {\frac {1}{12 \pi{r}^{2}} \left[ -16\,B\pi \,{r}^{2}+3\,a{r}^{2}+7
\,\rho_{{1}} \left( {\frac {13\,a}{210}}-{\frac {8\,B\pi }{45}}
 \right) {r}^{6}-\frac{3}{8}\,\rho_{{1}}{r}^{4} \right] },\label{eq17}
\end{equation}

Substituting Eqs.~(\ref{eq1}) and (\ref{eq6}) into Eq.~(\ref{eqE2}) we have
\begin{equation}
g_{tt}=C \frac{{{\mathrm e}^{\psi \left( r \right)
}}}{{r^{\frac{4}{3}}}\left(1-\frac{2m(r)}{r}\right)^\frac{1}{3} }, \label{eq18}
\end{equation}
where $\psi \left( r \right)=\frac{4}{3}\,\int \!{\frac {1-8\,B\pi \,{r}^{2}}{r-2\,m\left( r \right) }}
\,{\mathrm d}r$.

After evaluating $C$, based on the suitable boundary condition, Eq. (\ref{eq18}) can be written as
\begin{equation}
g_{tt}= {{\mathrm e}^{[\psi \left( r \right) -\psi \left( R \right)]}}{\frac{R^{\frac{4}{3}}}{r^{\frac{4}{3}}}}{\frac{\left(1-\frac{2M}{R}\right)^{\frac{4}{3}}}{\left(1-\frac{2m(r)}{r}\right)^\frac{1}{3}}}. \label{eq19}
\end{equation}

This is the {\it time-time} component of the interior metric of the ultra-dense spherical stellar system.

\begin{figure}[h]
\centering
\includegraphics[scale=.3]{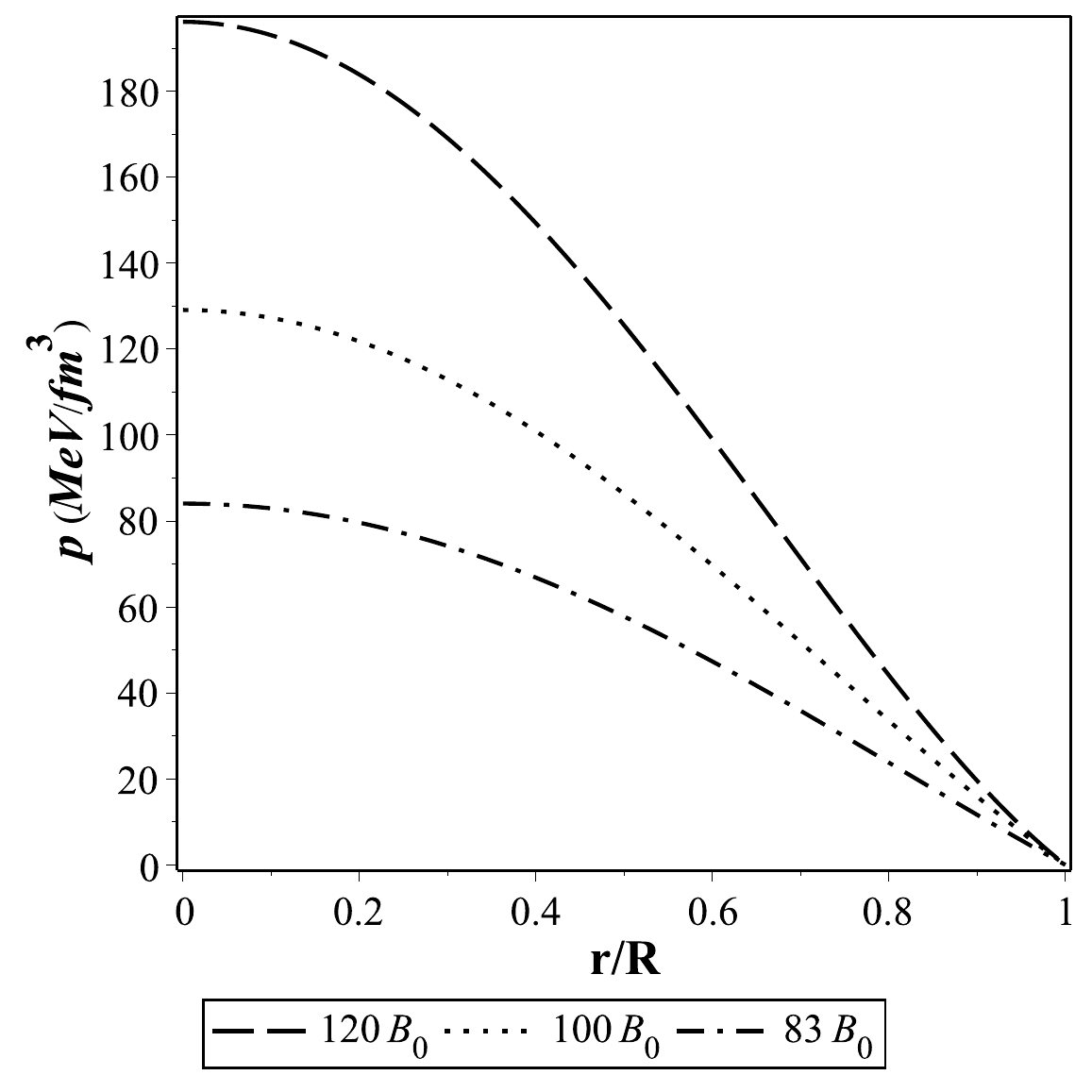}
\caption{Variation of the pressures as a function of the radial distance $r/R$ for the strange star $Cen~X-3$}
\label{Fig2}
\end{figure}

The nature of the pressure is shown in Fig. \ref{Fig2} which shows the physically acceptable feature. We have plotted 
the variation of the metric potentials $g_{tt}$ and $g_{rr}$ against the radial coordinate $r/R$ in Fig.~\ref{pot} 
which confirms that our system is free from the geometrical singularity.

\begin{figure}[h]
\centering
    {\includegraphics[scale=.3]{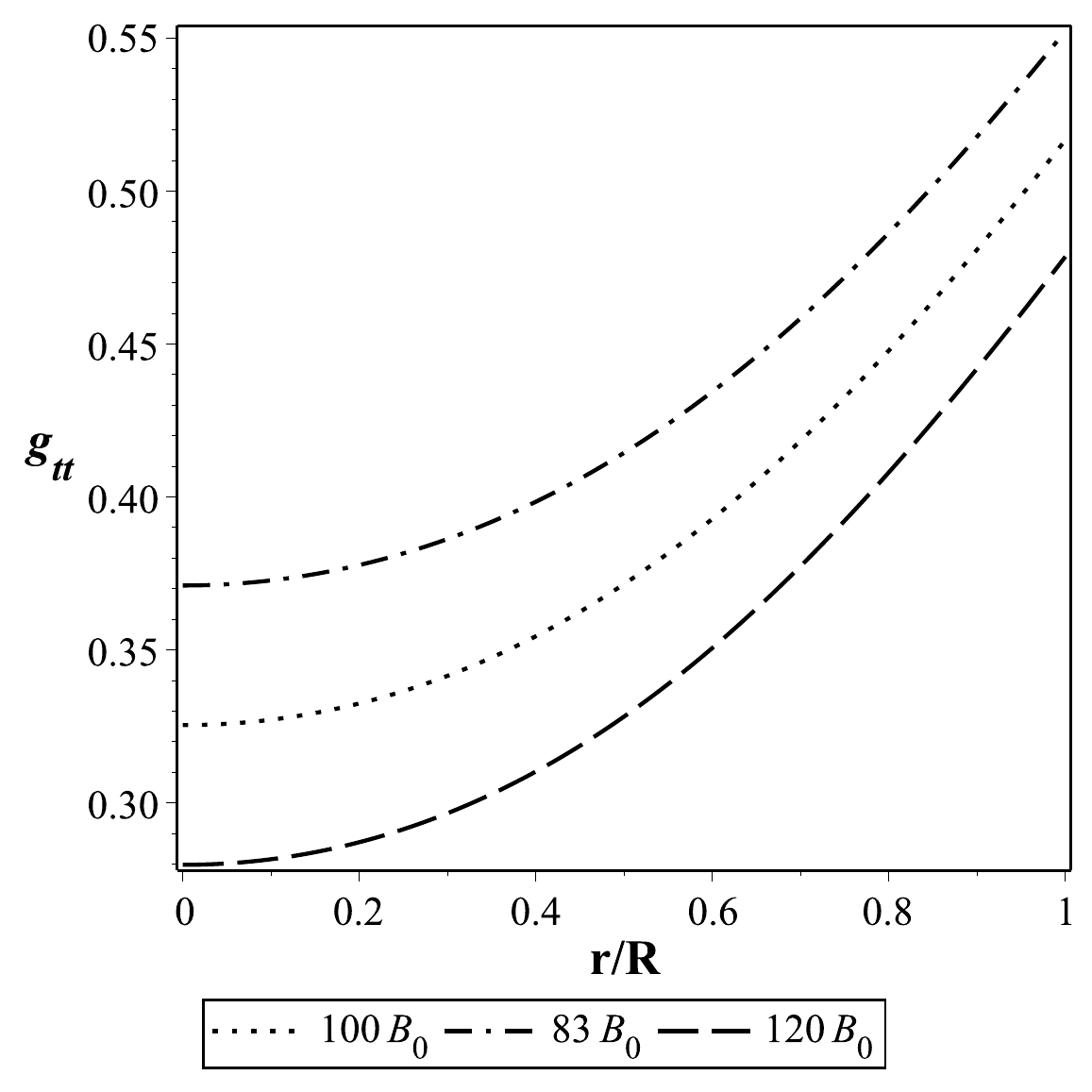}}
    {\includegraphics[scale=.3]{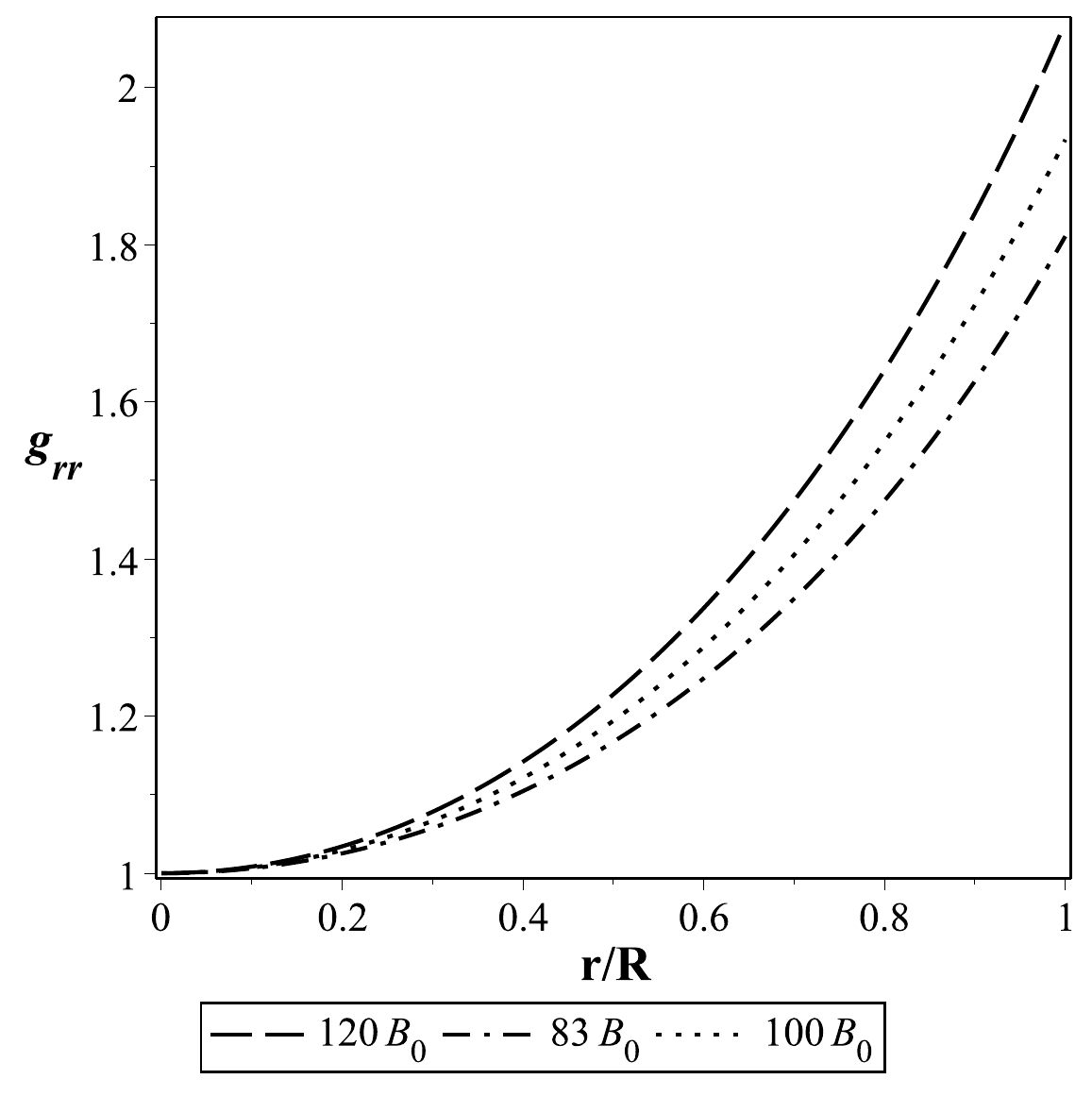}}
\caption{Variation of $g_{tt}$ and $g_{rr}$ as a function of the radial distance $r/R$ for the strange star candidate $Cen~X-3$}
\label{pot}
\end{figure}

\section{Physical properties of the stars}\label{physical}
In this section we are going to discuss different physical features
of the strange stars using the proposed model.

\subsection{Stability of the system}\label{stability}

\subsubsection{The Tolman-Oppenheimer-Volkoff (TOV) equation}\label{TOV}
To study the stability of the system we have checked the stability
equation given by Tolman~\cite{Tolman1939}, Oppenheimer and
Volkoff~\cite{Oppenheimer1939}. The TOV equation depicts the
equilibrium condition of a star subject to the gravitational
and hydrostatic forces. The generalized
TOV equation can be written as~\cite{Leon1993,Varela2010}
\begin{equation}
-\frac{M_g(\rho +p)}{r^{2}} e^{\frac{\lambda- \gamma }{2}}
-\frac{dp}{dr}=0, \label{eq23}
\end{equation}
where the effective gravitational mass $M_g$ of the system is defined as
\begin{equation}
M_g = \frac{1}{2}r^2e^{\frac{\gamma -\lambda}{2}} \gamma',
\label{eq24}
\end{equation}
with $\gamma \left( r \right)$ and $\lambda \left( r \right)$ are
respectively $\ln g_{{{\mathit tt}}}$ and $-\ln  \bigg[ 1-{\frac {2m
\left( r \right) }{r}} \bigg] $.

The TOV equation for our system can be translated as
\begin{equation}
\frac{2}{3}\,{\frac { \left(B -\rho\right) {g_{tt}}^{\prime}}{g_{tt}}} -\frac{1}{3} \frac{d
\rho}{dr} =0, \label{eq25}
\end{equation}
where the first term of the above equation is the gravitational force ($F_g$)
and the second term is the hydrostatic force ($F_h$) respectively, so that
for equilibrium of the system we should have
\begin{equation}
F_g + F_h =0. \label{eq26}
\end{equation}

We have drawn the forces in Fig. \ref{Fig3} which describes the overall
behavior of different forces.

\begin{figure*}[h]
\centering
    {\includegraphics[scale=.2]{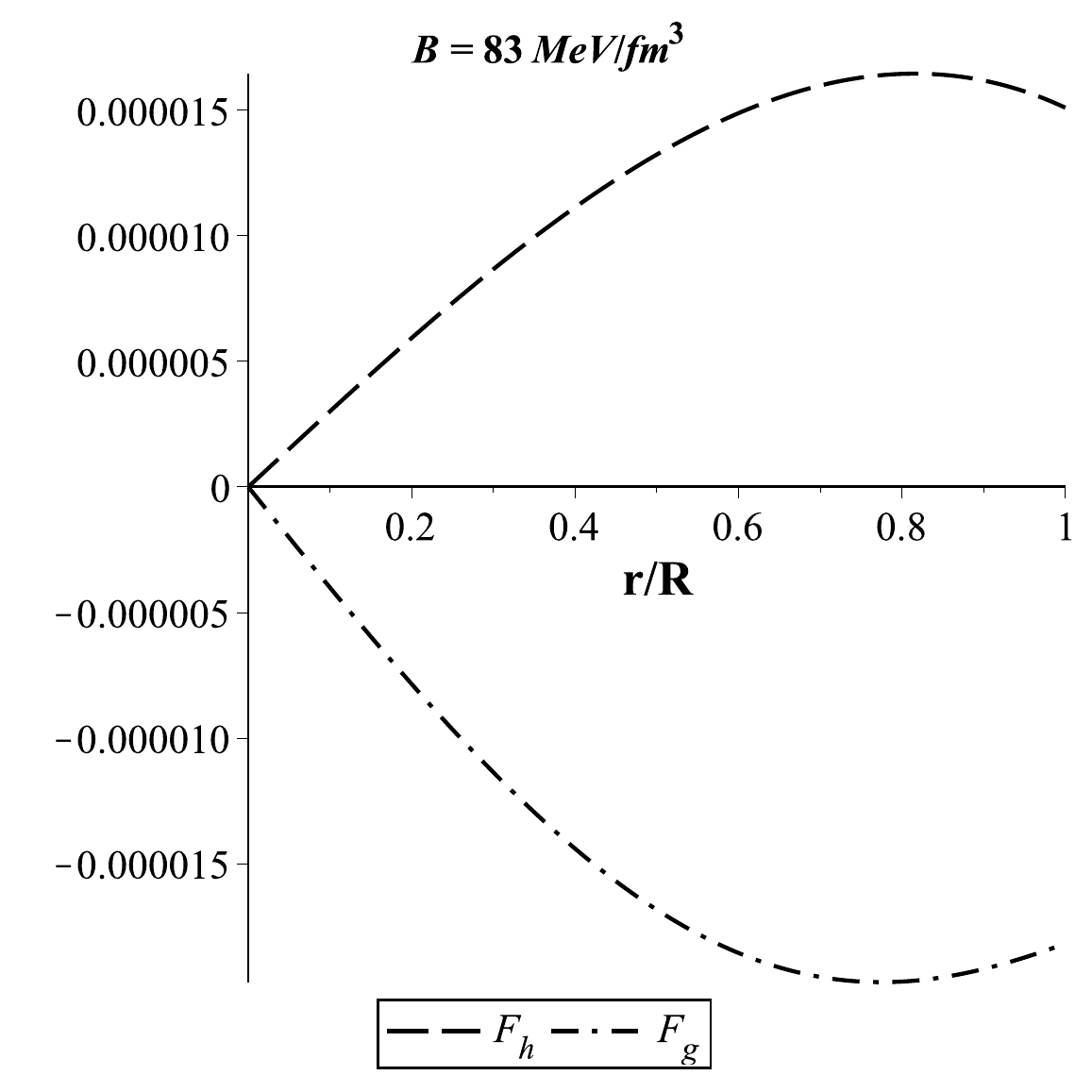}}
    {\includegraphics[scale=.2]{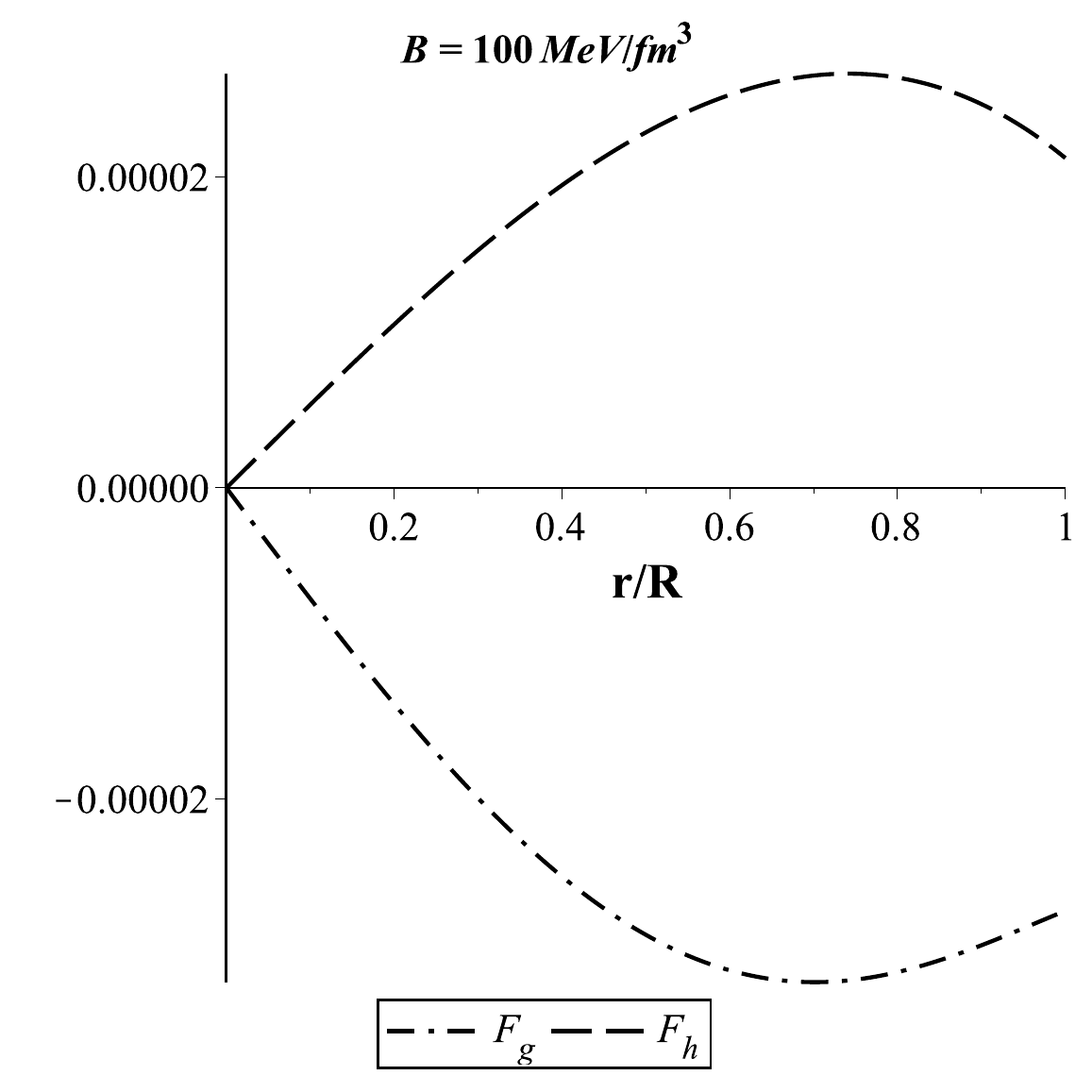}}
    {\includegraphics[scale=.2]{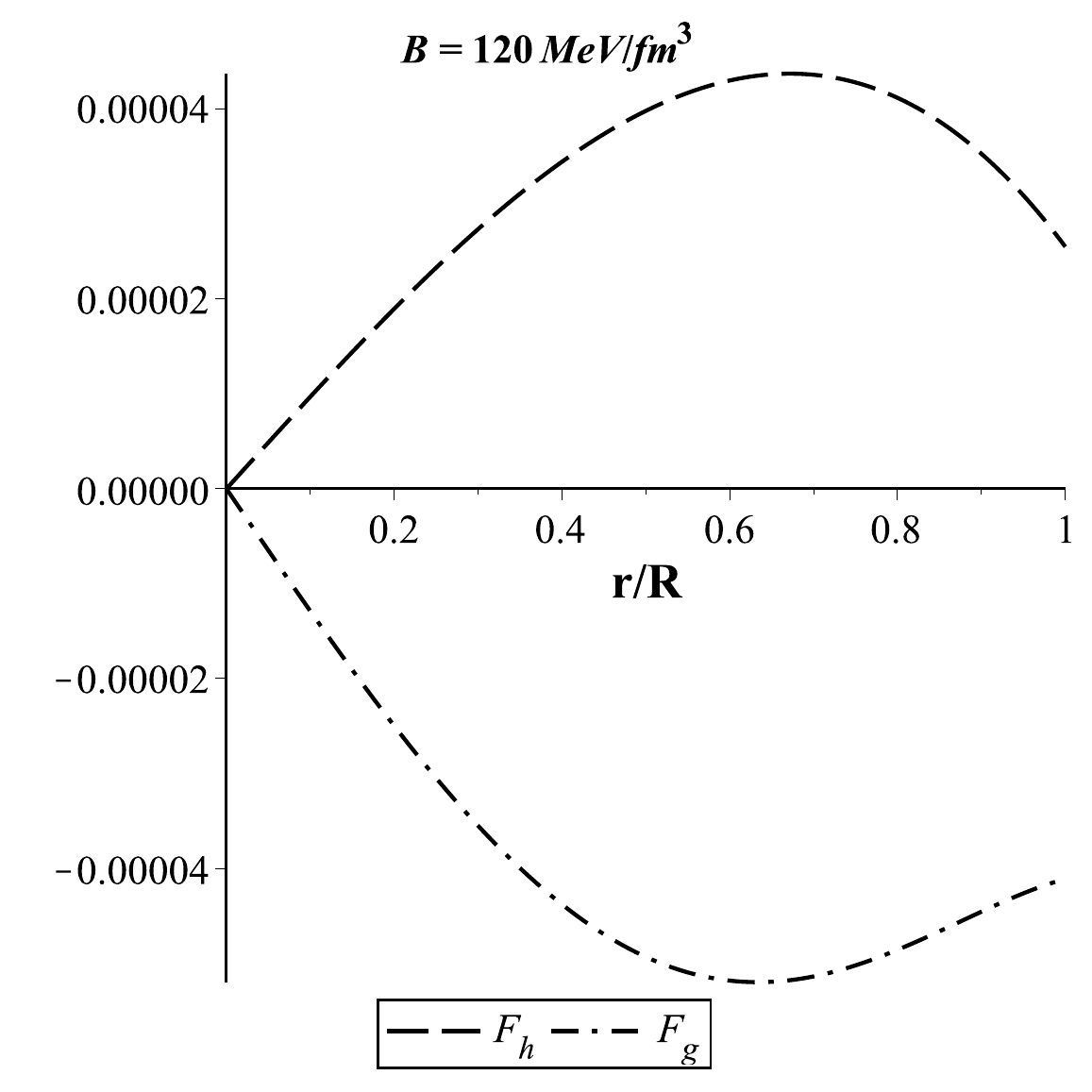}}
\caption{Variation of the different forces due to three values of bag constant, as a function of the radial distance $r/R$ for the strange star $Cen~X-3$}  
\label{Fig3}
\end{figure*}

\subsubsection{Adiabatic Index}\label{adia_ind}
For the isotropic spherical stellar system, Chandrasekhar~\cite{Chandrasekhar1964} in his pioneering works has shown that the essential and sufficient condition for the stability against the radial pulsation is the adiabatic index ($\Gamma$) of the system should be greater that $4/3$, i.e., $\Gamma > 4/3$. From our model, we have

{\small{\begin{equation}
\Gamma=\frac{\rho+p}{p} \frac{dp}{d\rho}={\frac { \left[  \left( 1792\,B\pi -624\,a \right) {r}^{4}+540\,{r}^{2
} \right] \rho_{{1}}+5760\,B\pi -4320\,a}{ \left[  \left( 1344\,B\pi -
468\,a \right) {r}^{4}+405\,{r}^{2} \right] \rho_{{1}}+17280\,B\pi -
3240\,a}}.\label{eq26a}
\end{equation}}}

\begin{figure*}[h]
\centering
\includegraphics[scale=.3]{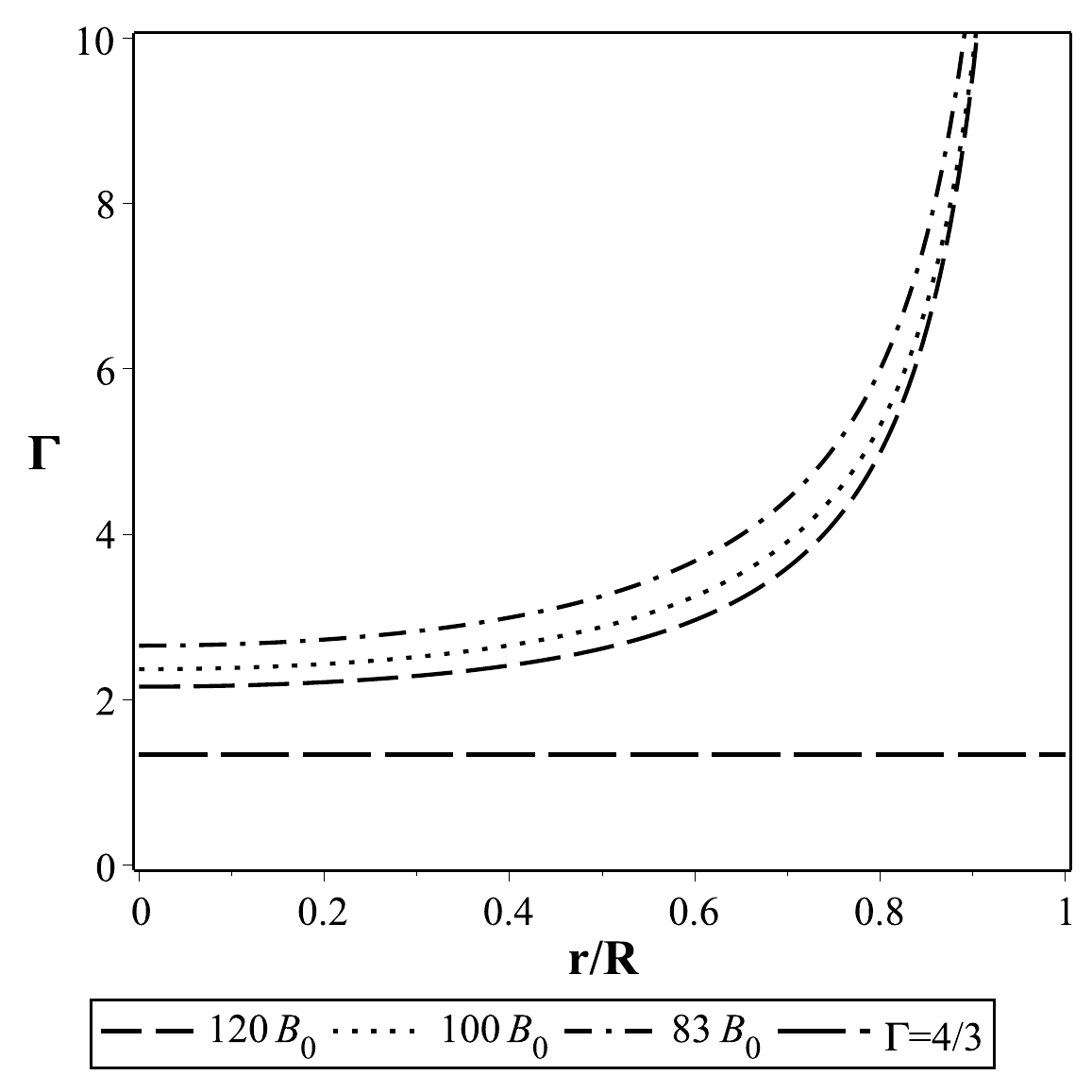}
\caption{Variation of the adiabatic index as a function of the radial distance $r/R$ for the strange star $Cen~X-3$} 
\label{Fig3a}
\end{figure*}

From Fig. \ref{Fig3a} it is clear that adiabatic index for our system is greater than $4/3$
in all the interior points of the system, which confirms that the system is stable by nature.

\subsection{Energy conditions}\label{energ_con}
The ultra-dense spherically symmetric system should satisfy all the
energy conditions, viz. null energy condition (NEC), weak energy
condition (WEC), strong energy condition (SEC) and dominant energy condition (DEC) respectively
given by
\begin{eqnarray}\label{ECON1}
& \qquad NEC: \rho+p \geq 0,\\\label{ECON2}
& \qquad WEC: \rho+p \geq 0,~\rho \geq 0,\\\label{ECON3}
& \qquad SEC: \rho+p \geq 0,~\rho+3p \geq 0, \\\label{ECON3}
& \qquad DEC: \rho \geq 0;~\rho - p \geq 0.
\end{eqnarray}

In Fig. \ref{Fig5} we have shown the behavior of all the above mentioned 
energy inequalities and it is clear that our system is consistent with all the energy conditions.

\begin{figure*}[h]
\centering
{\includegraphics[scale=.2]{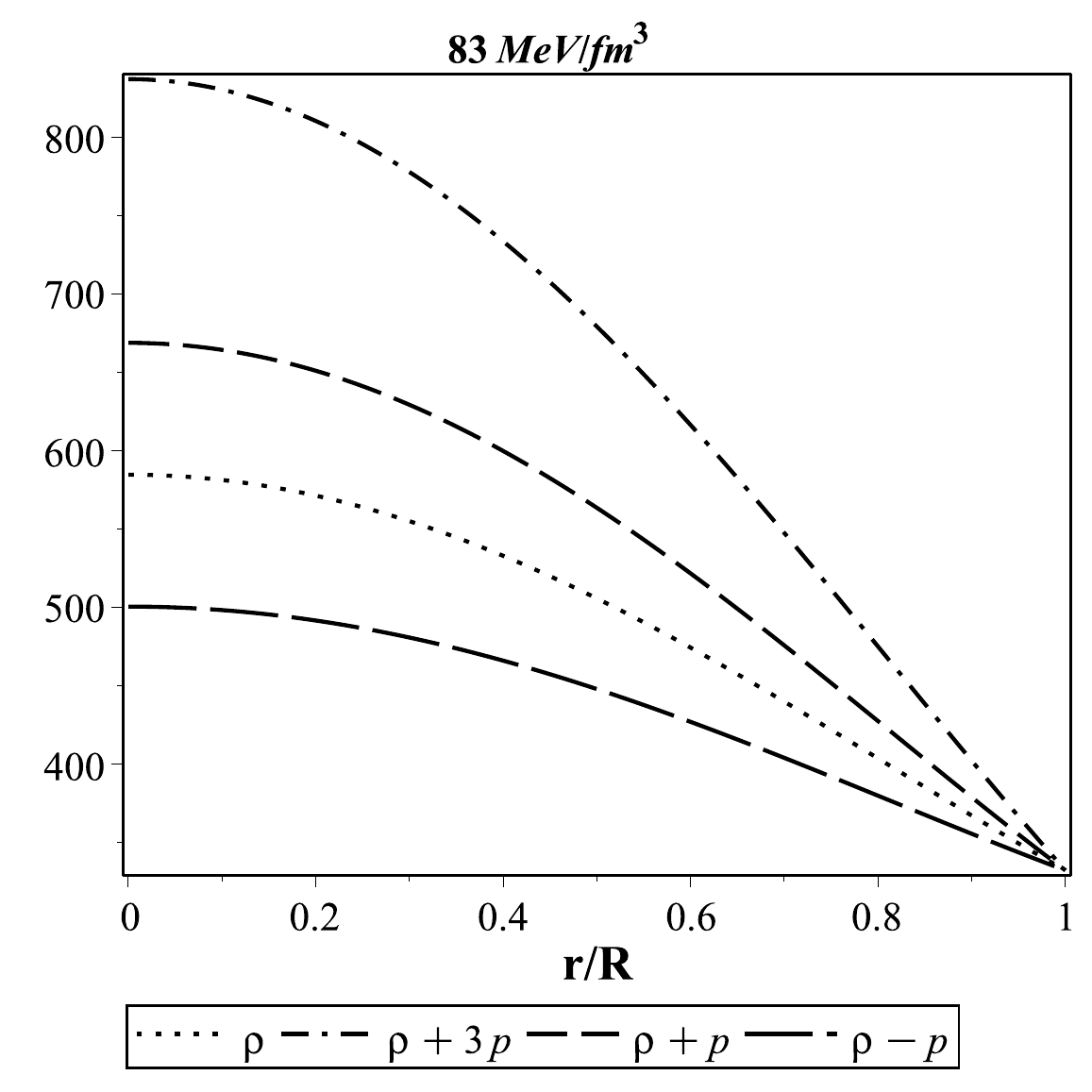}}
{\includegraphics[scale=.2]{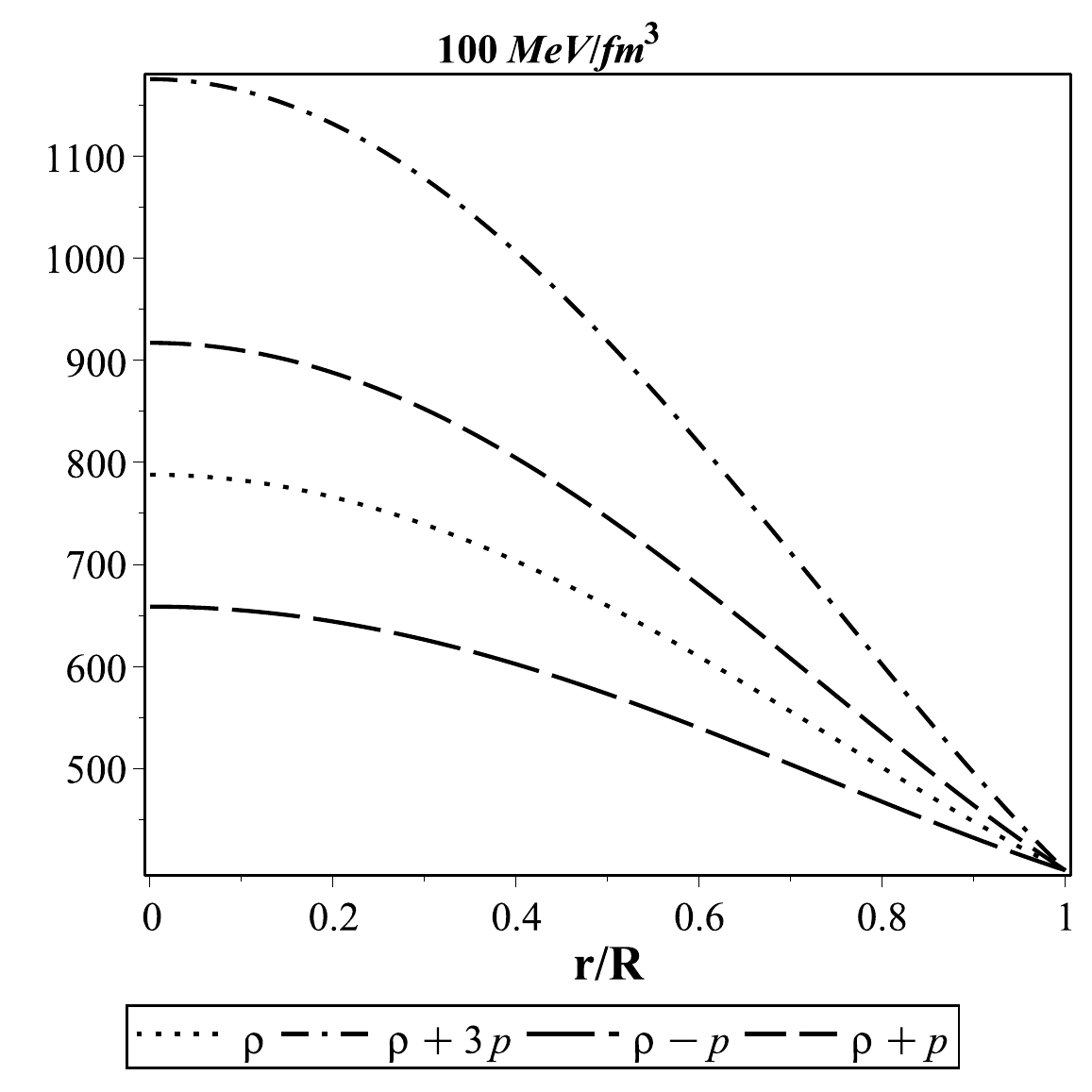}}
{\includegraphics[scale=.2]{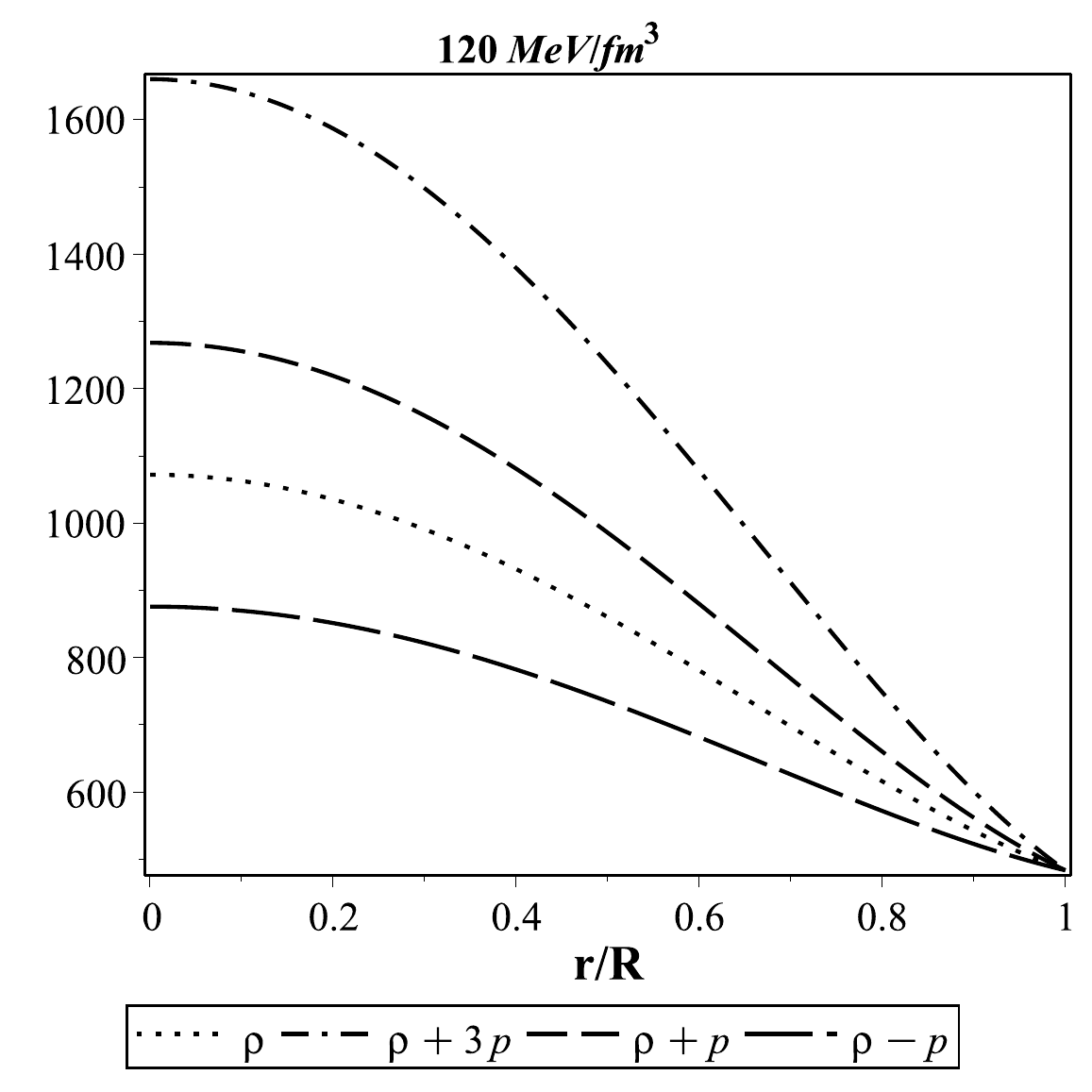}}
\caption{Variation of the different energy conditions as a function of the radial distance $r/R$ for the strange star $Cen~X-3$}  
\label{Fig5}
\end{figure*}

\subsection{Surface Redshift}\label{reds.}
The compactification factor of a star is defined as the mass-to-radius ratio
of the system, i.e. $u(r) =  m(r)/r $. According to the
condition of Buchdahl~\cite{Buchdahl1959} the maximum allowed mass
radius ratio is $\leq $ 8/9~($\approx$ 0.89) for the perfect fluid sphere.

For our system the compactification factor is
\begin{eqnarray}
u(r) =a{r}^{2}-{\frac {3\,\rho_{{1}}{r}^{4}}{40}}+\rho_{{1}} \left( {\frac {13\,a}{210}}-{\frac {8\,B\pi }{45}}
 \right) {r}^{6}.
\label{eq27}
\end{eqnarray}

Surface redshift $(Z_s)$ of a star is defined as
\begin{equation}
1+Z_s = \left[1-2u(R)\right]^{- \frac{1}{2}}, \label{eq28}
\end{equation}
which for the above studied system is given by
\begin{eqnarray}
Z_s={\frac {1}{\sqrt {1-2\,a R^2 +\frac {3\,\rho_{1}R^4}{20} 
- 2\,\rho_{1} \left(\frac {13\,a}{210}-\frac {8\,B\pi }{45} \right) R^6 }}} - 1. \label{eq29}
\end{eqnarray}

\begin{figure*}[h]
\centering
{\includegraphics[scale=.3]{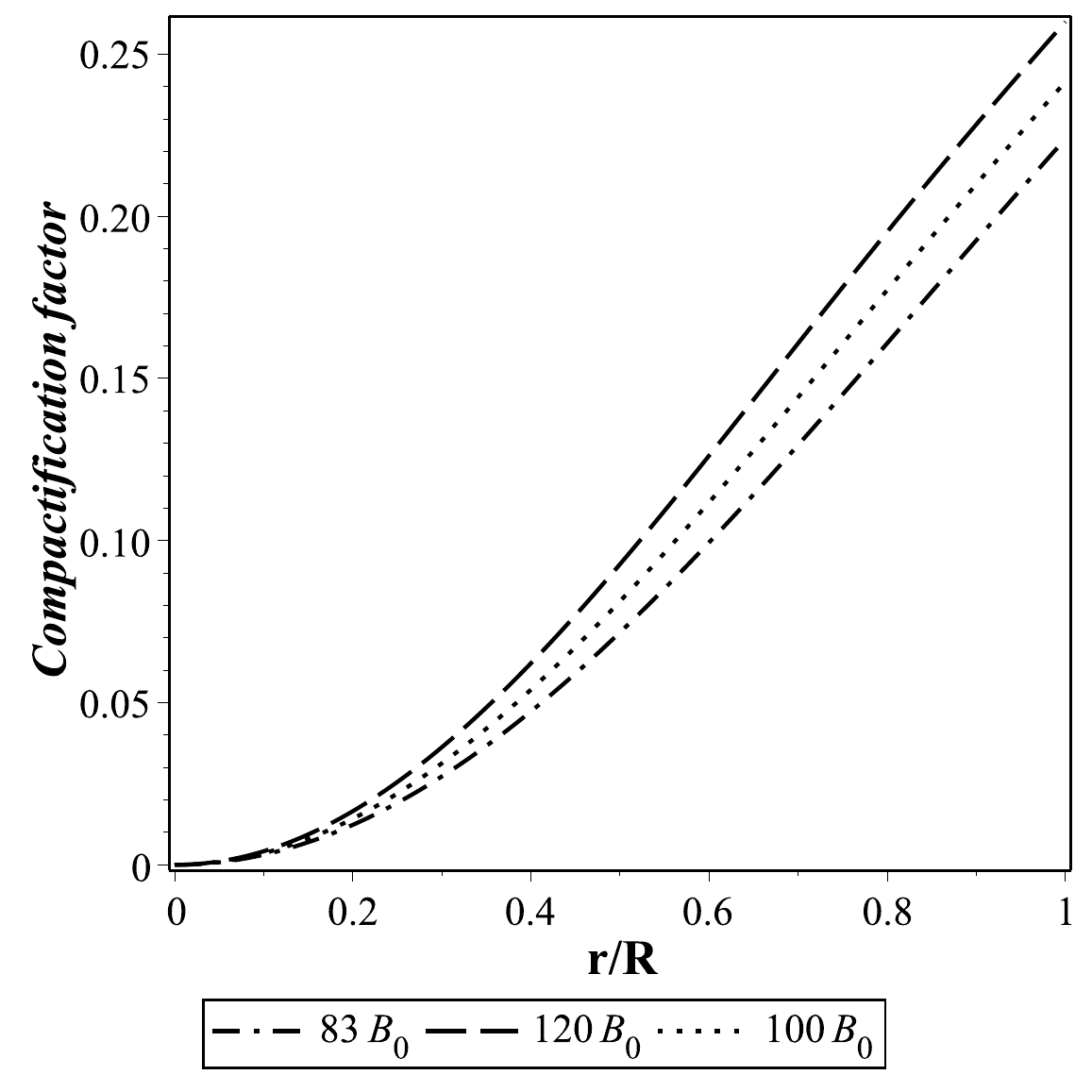}}
{\includegraphics[scale=.3]{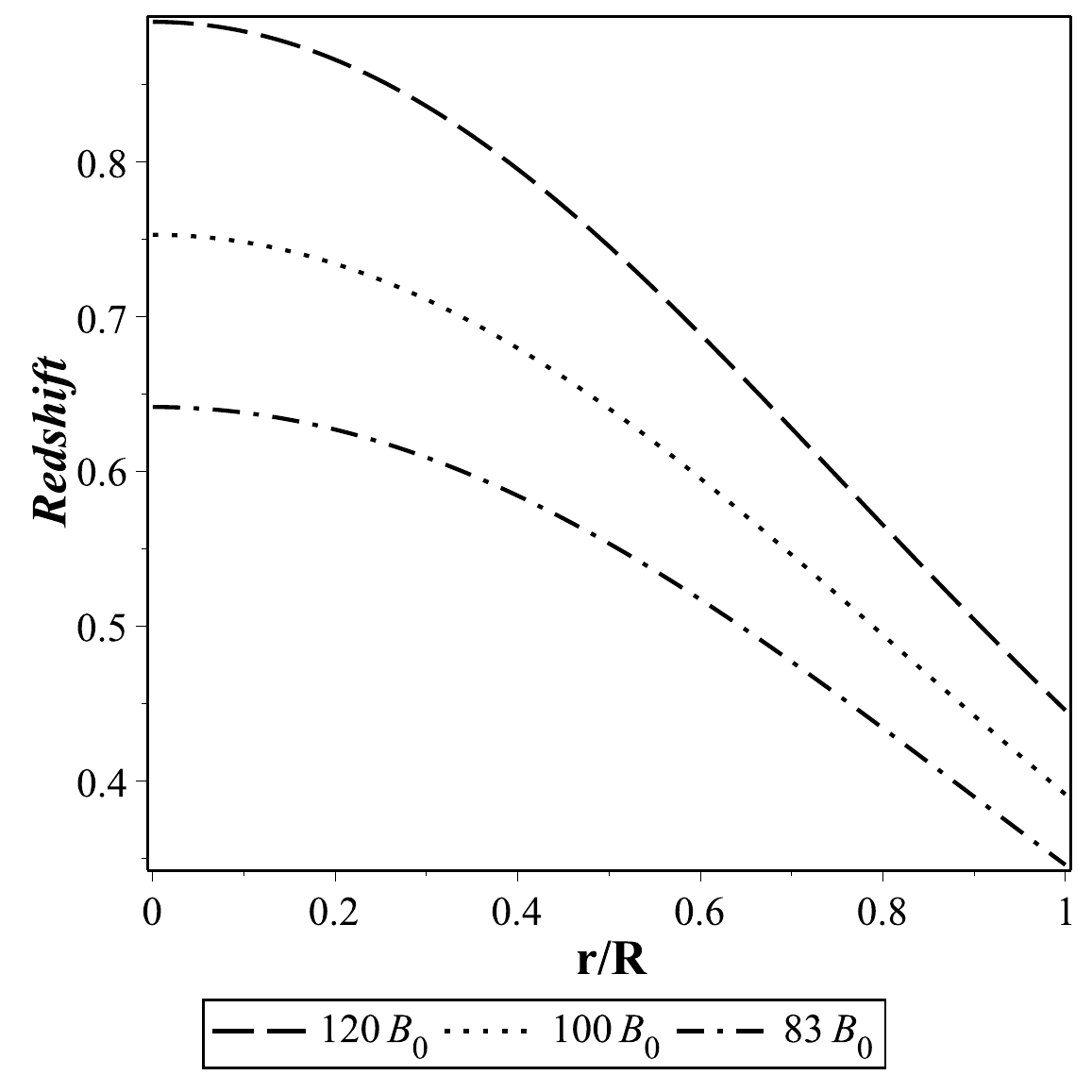}}
\caption{Variation of the compactness (left panel) and redshift (right panel) as a function of the radial distance $r/R$ for the strange star $Cen~X-3$}  
\label{Fig6}
\end{figure*}

Variation of the compactification factor with respect to the 
fractional radial coordinate $r/R$ are shown in the left panel of Fig. \ref{Fig6}. 
Further, we have shown variation of the redshift function, 
$Z=\left(\frac{1}{\sqrt{g_{tt}}}-1\right)$ with the radial coordinate $r/R$ 
in the right panel of Fig.~\ref{Fig6}. From the X-ray spectrum of the stars, the 
surface redshift $Z_s$ can be easily observed and correspondingly compactness can 
be calculated.

\section{A comparative study}\label{com_st}
To study the physical properties of the system we choose the star
$Cen~X-3$ as a representative of the strange stars, having parameters $a= 2448.995 $ $MeV/{fm}^3$, $R=9.819$~km
and mass $m(R) = 1.49~{{M}_{\odot}}$ for $B=83~MeV/{fm}^3$. 

With the help of the chosen values of radius and mass, we have shown different physical properties of the proposed structure of strange stars (Table~\ref{table1}). The observed mass in Table \ref{table1} is available in the literature~\cite{Rawls2011,Guver2010a,Freire2011,Guver2010b,Demorest2010}. However, in the lower as well as higher mass limits we do not yet find any observed stars whose mass tally with our prepared data sheet and thus kept blank.

\afterpage{
\begin{landscape}
 

\centering
\captionof{table}{Physical parameters of the different observed strange star candidates for $B=83~MeV/{fm}^3$}
\begin{tabular}{cccccccccccccccccccccccccc}
\hhline{=========} 
Radius & Predicted Mass & $a$ & ${\rho}_c$ & ${p}_c$ & $\frac{2M}{R}$ & $Z_s$  &  Observed Stars\\ 
(Km) & ($M_\odot$) & $(MeV/{fm}^3)$ & ($gm/{{cm}^3}$) & ($dyne/{{cm}^2}$) & &  & \\
\hline
9.6 & 1.12 & $2220.039$ & $9.448\times {{10}^{14}}$ & $1.055\times {{10}^{35}}$ & 0.41 & 0.30 & - \\ 
\ 
9.7 & 1.17 & $2308.448$ & $9.824\times {{10}^{14}}$ & $1.168\times {{10}^{35}}$ & 0.43 & 0.33 & - \\ 
\ 
9.819 & 1.49 & $2448.995$ & $1.042\times {{10}^{15}}$ & $1.347\times {{10}^{35}}$ & 0.45  & 0.35 & $  Cen~X-3$~\cite{Rawls2011} \\ 
\ 
9.92 & 1.58 & $2614.478$ & $1.113\times {{10}^{15}}$ & $1.558\times {{10}^{35}}$ & 0.47  & 0.37 & $4U~1820-30$~\cite{Guver2010a} \\ 
\ 
10.017 & 1.667 & $2814.720$ & $1.198\times {{10}^{15}}$ & $1.813\times {{10}^{35}}$ & 0.49  & 0.40 & $PSR~J1903+327$~\cite{Freire2011} \\ 
\ 
10.105 & 1.74 & $2988.514$ & $1.272\times {{10}^{15}}$ & $2.035\times {{10}^{35}}$ & 0.51  & 0.43 & $4U~1608-52$~\cite{Guver2010b} \\ 
\ 
10.143 & 1.77 & $3051.987$ & $1.299\times {{10}^{15}}$ & $2.116\times {{10}^{35}}$ & 0.52  & 0.44 & $Vela~X-1 $~\cite{Rawls2011} \\ 
\ 
10.2 & 1.815 & $3131.328$ & $1.333\times {{10}^{15}}$ & $2.217\times {{10}^{35}}$ & 0.53  & 0.46 &  - \\ 
\ 
10.3 & 1.879 & $3234.850$ & $1.377\times {{10}^{15}}$ & $2.349\times {{10}^{35}}$ & 0.54  & 0.474 & -\\ 
\ 
10.465 & 1.97 & $3339.127$ & $1.421\times {{10}^{15}}$ & $2.482\times {{10}^{35}}$ & 0.56  & 0.51  & $PSR~J1614-2230$~\cite{Demorest2010}\\ 
\hhline{=========}
\end{tabular}\label{table1}

\bigskip 
\bigskip 
\bigskip 
\bigskip

\centering
\captionof{table}{Physical parameters of the strange star candidate $Cen~X-3$, having mass $1.49~{M_{\odot}}$~\cite{Rawls2011} due to different values of $B$}
\begin{tabular}{cccccccccccccccccccccccccc}
\hhline{=========} 
B & Radius  & $a$ & ${\rho}_c$ & Surface Density & $p_c$ & $2M/R$ & $Z_s$  \\ 
$(MeV/{fm}^3)$ & $(Km)$ & $(MeV/{fm}^3)$ & $(gm/{cm}^3)$ & $(gm/{cm}^3)$ & $(dyne/{cm}^2)$ & & \\
\hline
83 & 9.819 & 2448.995 & $1.042\times {10}^{15}$ & $5.927\times {10}^{14}$ & $1.347\times {10}^{35}$ & 0.45 & 0.35 \\ 
\ 
100 & 9.095 & 3299.834 & $1.404\times {10}^{15}$ & $7.139\times {10}^{14}$ & $2.068\times {10}^{35}$ & 0.48 & 0.39 \\ 
\ 
120 & 8.43 & 4490.706 & $ 1.911\times{10}^{15}$ & $8.621\times {10}^{14}$ & $ 3.143\times{10}^{35}$ & 0.52 & 0.44 \\ 
\hhline{=========}
\end{tabular}\label{table2}

 
\end{landscape} 
}

In Table~\ref{table2} we have presented a data sheet for different physical parameters of the strange star candidate $Cen~X-3$ due to three chosen values of $B$ as $83~MeV/{fm}^3$, $100~MeV/{fm}^3$, and $120~MeV/{fm}^3$. We find that as the values of $B$ increase the stellar system becomes more compact and energy density within the stars increases gradually. With the increasing values of $B$ the observed value of the mass of $Cen~X-3$~\cite{Rawls2011} is achieved for the gradually decreasing values of radius, i.e., the stellar system does shrink. The values of surface redshift also rise with the increasing values of $B$.

\section{Discussions and conclusions}\label{discsn}
In this article, we have tried to solve stellar hydrodynamic equation (i.e. TOV equation) directly 
by using HPM technique and derived the mass profile for the spherically symmetric compact stellar 
system. Further, we have obtained expressions for different physical parameters, viz., $g_{tt}$,
$g_{rr}$, $\rho$ and $p$. The salient features of the proposed stellar model from the present 
investigation are as follows:

(1) Our model is compatible with the compact stars, especially that of strange stars as seen from the
comparative study of the previous Sec.~\ref{com_st}.

(2) In Sec.~\ref{stability}, by studying different tests, viz., equilibrium of different forces and stability against radial pulsation, we find out that our model predicts a completely stable stellar system. Also, to be consistent with the causality condition the square of the sound speed $(v_s^2)$ must lie within the limit $0$ to $1$. In our work, for the specified sets of data, we find that ${v_s^2}=\frac{d\,p}{d\rho}=\frac{1}{3}$, i.e. $0 \leq {v_s^2} \leq 1$, which also confirms the stability of the system.

(3) Figs.~\ref{Fig1},~\ref{Fig2} and \ref{pot} show interesting features that the physical parameters, viz., $\rho$, $p$, $g_{tt}$ and $g_{rr}$ have finite values at the center, which confirm that our system is completely free from any sort of geometric or physical singularities.

(4) From our model, we find that $\frac{2\,M}{R} < \frac{8}{9}$ for all the strange star candidates.
Hence, Buchdahl condition~\cite{Buchdahl1959} holds good for our system. Also, as $r \rightarrow 0$
we find $m(r) \rightarrow 0$ which shows that the mass function is regular at the center.

(5) In the present paper, with the help of the chosen radius and specific 
value of the bag constant~\cite{Rahaman2015} we have derived the value of the mass 
for different possible strange star candidates (shown in Table~\ref{table1}), 
whereas in Table~\ref{table2} we have shown the possible variation of the physical 
parameters for different chosen values of bag constants. However, it is worth 
mentioning that the values of $B$ are chosen randomly to present the numerical and 
graphical outputs of the solutions.

6) Using the chosen numerical values of the radius and bag constant, we have calculated 
different properties of the interior solution of the spherical symmetric body and also 
graphically presented different physical features of the model. From Fig. \ref{Fig5} it 
is clear that our model satisfies all the energy conditions which is an essential condition 
for a compact stellar system to be physically valid. In the present investigation, we 
find high surface redshift values ($0.30-0.51$), which are quite relevant for strange star candidates. 

So both the data, redshift as well as mass, indicate that the model studied in the present paper is a 
representative of a compact star and is suitable to explore different properties of strange stars.

\section*{Acknowledgements}
SR and FR are thankful to the Inter-University Centre for Astronomy and Astrophysics (IUCAA), 
Pune, India for providing Visiting Associateship under which a part of this work was carried 
out. SR is also thankful to the authority of The Institute of Mathematical Sciences (IMSc), Chennai, 
India for providing all types of working facility and hospitality under the Associateship 
scheme. FR is also grateful to DST-SERB (EMR/2016/000193), Govt. of India for providing 
financial support. A part of this work was completed while DD was visiting IUCAA and the 
author gratefully acknowledges the warm hospitality and facilities at the library there. 
We all are thankful to the anonymous referee for several pertinent comments which have helped us 
to upgrade the manuscript substantially.\\

{\footnotesize{}}
\end{document}